\documentclass[]{spie}  

\usepackage{amsmath,amsfonts,amssymb}
\usepackage{graphicx}
\usepackage[colorlinks=true, allcolors=blue]{hyperref}

\title{MIRC-X/CHARA: sensitivity improvements with an ultra-low noise SAPHIRA detector}

\author[a]{Narsireddy Anugu}
\author[b,c]{Jean-Baptiste Le Bouquin}
\author[b]{John D. Monnier}
\author[a]{Stefan Kraus}
\author[b]{Jacob Ennis}
\author[c]{Cyprien Lanthermann}
\author[b]{Benjamin R. Setterholm}
\author[a]{Claire L. Davies}
\author[d]{Theo ten Brummelaar}
\author[b]{Mariam Haidar} 
\author[b]{Veronika Dubravec} 
\author[b]{Scott Peters}

\affil[a]{School of Physics and Astronomy, University of Exeter, Exeter EX4 4QL, UK}
\affil[b]{University of Michigan, Ann Arbor, MI 48109, USA}
\affil[c]{Institut de Planetologie et d'Astrophysique de Grenoble, Grenoble 38058, France}
\affil[d]{CHARA Array, Georgia State University, Atlanta, GA 30302, USA}

\authorinfo{Further author information: (Send correspondence to N. A.)\\N. A.: E-mail: n.anugu@exeter.ac.uk, Telephone: +1 734 263 4911}

\begin{document} 
\maketitle

\begin{abstract}
MIRC-X is an upgrade of the six-telescope infrared beam combiner at the CHARA telescope array, the world's largest baseline interferometer in the optical/infrared, located at the Mount Wilson Observatory in Los Angeles. The upgraded instrument features an ultra-low noise and fast frame rate infrared camera (SAPHIRA detector) based on e-APD technology. We report the MIRC-X sensitivity upgrade work and first light results in detail focusing on the detector characteristics and software architecture.
\end{abstract}


\section{Introduction}

The Michigan Infrared Combiner (MIRC) is the six-telescope infrared beam combiner at the CHARA telescope array, the world's largest baseline interferometer in the optical/infrared, located at the Mount Wilson Observatory in Los Angeles. MIRC has been operating at the CHARA array since 2005. A detailed description of MIRC can be found elsewhere\cite{mirc2004,Monnier2010}(e.g., Monnier et al. 2004, 2010). In short, it operates primarily at H-band and uses single-mode fibres to spatially filter out distorted wave fronts due to atmospheric turbulence. The six-telescope beams are arranged in a line with non-redundant spacing between them. The corresponding fringes for the 15 pairs are disentangled in the image plane by simple Fourier analysis. It provides $\lambda/{2B_\mathrm{max}} \sim 0.5$ milliarcseconds angular resolution with 15 baselines and 20 closure triangles. 

MIRC has produced landmark results in various areas of stellar astrophysics, including imaging of stellar surface structures of rapid rotators \cite{Monnier2007},  eclipsing system $\epsilon$ Aurigae\cite{Kloppenborg2010}, and spot networks on magnetically active stars\cite{Roettenbacher2016}. Based on its success, Exeter and Michigan University groups upgraded the sensitivity of MIRC with the aim to image faint protoplanetary disks, which were not observable before.

The upgrade of MIRC (MIRC-X after upgrade) is planned in two phases. Phase 1 includes the commissioning of an ultra-fast, ultra low read noise near-infrared camera and changes in the MIRC control software architecture. We have equipped MIRC with a C-RED ONE camera\cite{credone2016} from First Light Imaging. The camera uses a SAPHIRA detector\cite{finger2014}, which provides  10-30 times lower read noise than the previously-installed PICNIC detector. It also allows MIRC-X to act as fringe tracker for other science beam combiners at CHARA.  In Phase 2, upgrade of optics and fibers are planned to: (1) address the over sampling of fringes, which is caused by the pixel scale difference between the old PICNIC (40\,$\mu$m) and the new C-RED ONE (24\,$\mu$m) cameras; (2) improve the transmission in H-band; (3) enable simultaneous J and H-band observations; and (4) minimize the cross-talk between different baselines. Please refer Kraus et al. in these proceedings~\cite{Kraus2018} for more details.

The science cases of MIRC-X, in order of priority, are to obtain (a) imaging of the hot dust in young stellar objects (T-Tauri and Herbig Ae/Be stars), 
(b) single-field astrometry to enable the detection of Hot Jupiter planets in binary systems, and (c) polarimetry in Stokes Q and U parameters. Together with MYSTIC  (Monnier et al.\cite{Monnier2018} in these proceedings) co-phasing, we also plan to enable observations in spectral lines, such as He-I, Pa-$\gamma$ and Pa-$\beta$ (see Kraus et al.\cite{Kraus2018} in these proceedings). MYSTIC is a six-telescope K-band beam combiner, planned for commissioning in 2019.

This paper focuses on the Phase 1 upgrade of MIRC in terms of instrument sensitivity and software architecture.  The second section describes the C-RED ONE camera and associated hardware installation. The third section details the software development including the new data acquisition and upgrade of CHARA-compliant servers and Graphical User Interfaces (GUIs). This follows the installation of polarization controllers. In the next section, MIRC-X performance is presented as a function of avalanche gain and readout mode configurations, including comparison with the previous PICNIC detector.  Next, on-sky performance of MIRC-X is reported. In the end, the plan of Phase 2 work is outlined.

\section{Installation of C-RED ONE for MIRC}
\begin{figure} [h!]
	\begin{center}
		\begin{tabular}{c} 
			\includegraphics[width=5in]{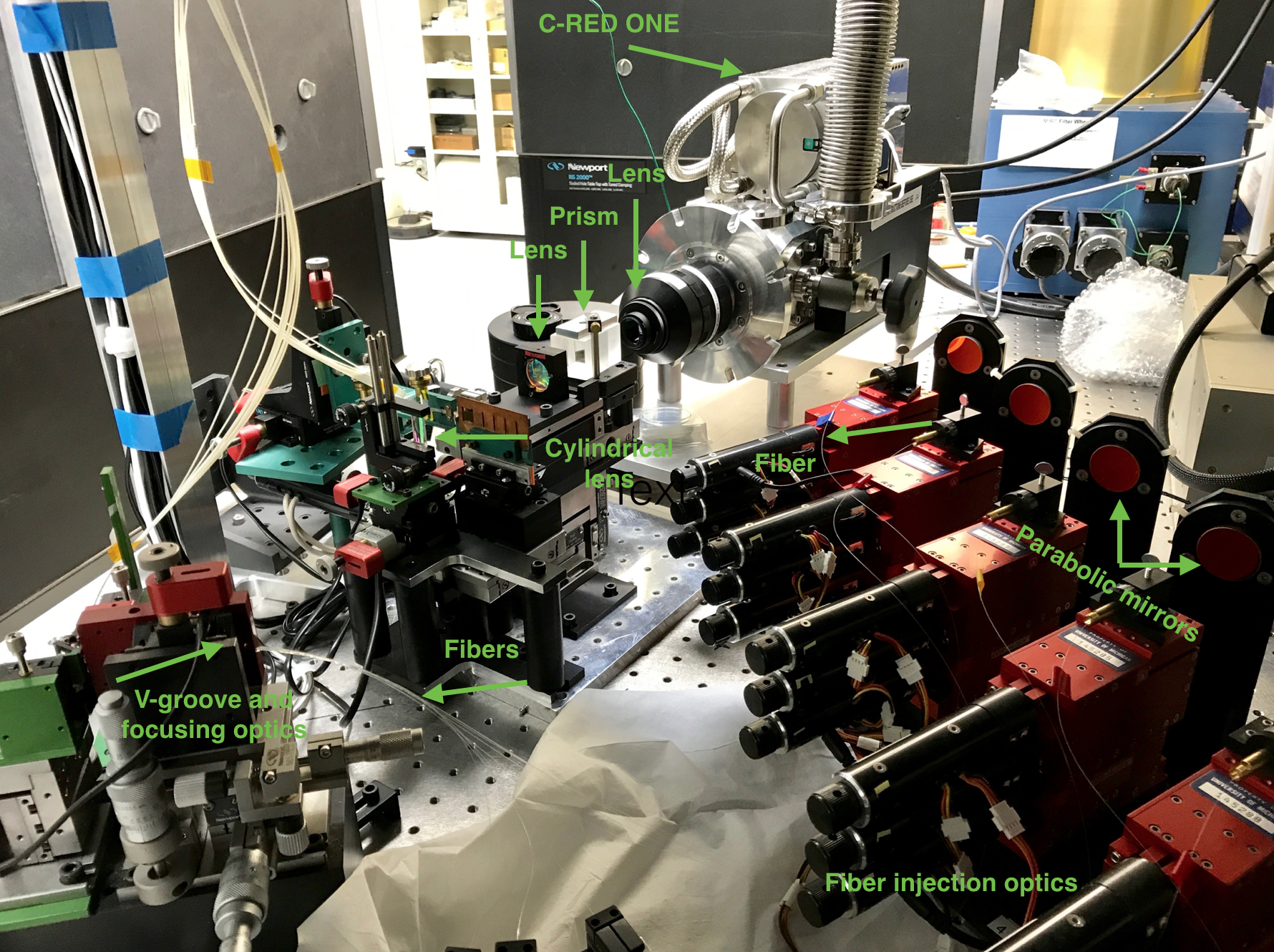}\\
			\includegraphics[width=5in]{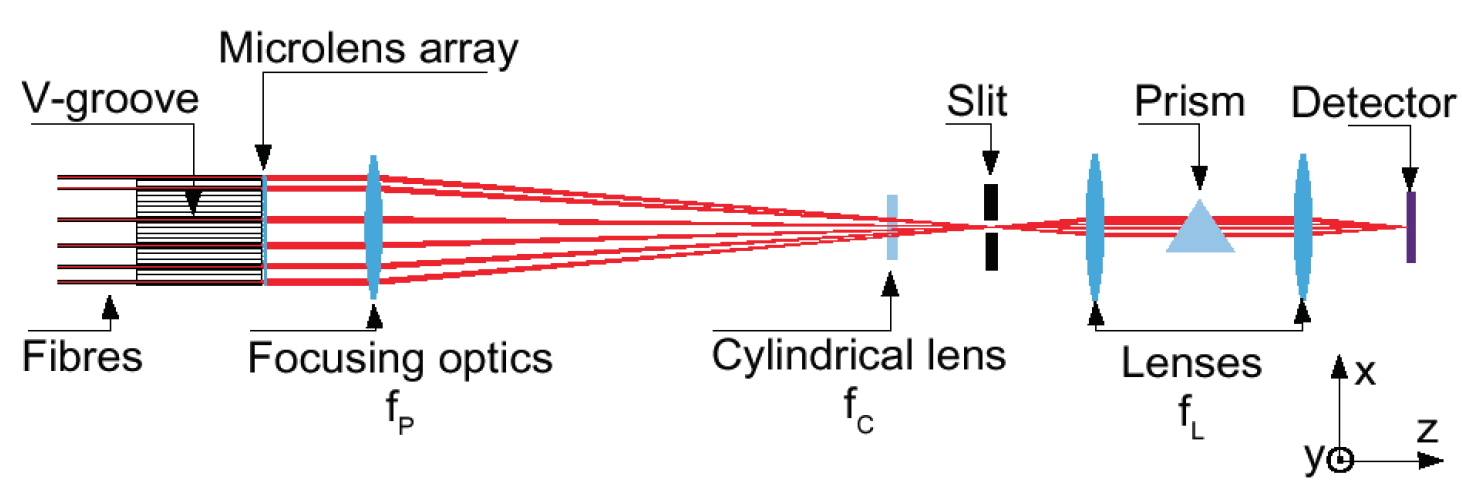}
		\end{tabular}
	\end{center}
	\caption[example] 
	{ \label{fig:MIRCX} 
		(Top) MIRC-X: Installation of C-RED ONE at MIRC optics. The parabolic mirrors inject light into single mode fibers and these fibers are arranged into a non-redundant pattern in a v-groove array. Then a spherical mirror focuses these six-beams to produce interference fringes at the input slit of the camera-spectrometer. (Bottom) Optical layout of MIRC (Figure credit: Monnier et al. 2010\cite{Monnier2010}).  
	}
\end{figure}

\subsection{C-RED ONE camera}

\subsubsection{Overview}
The C-RED ONE camera from First Light Imaging (Figure~\ref{fig:MIRCX}, Gach et al. 2016\cite{credone2016} and references within) is capable of capturing up to 3500 full frames per second with a sub-electron readout noise. It uses Mark13 model SAPHIRA detector developed by Leonardo, UK. SAPHIRA has $320\times256$ pixel HgCdTe linear electron avalanche photodiode (e-APD) arrays and a $24~\mu \mathrm{m}$ pixel pitch.  SAPHIRA is sensitive to $0.8-3.5~\mu \mathrm{m}$ wavelengths and is transparent at longer wavelengths. It amplifies the signal from incident individual photons by the mean of avalanche gains while keeping the read noise constant. At avalanche gains higher than 20, the effective readout noise is typically $<1\,e^{-}$ (sub-electron regime).

SAPHIRA has 32 parallel video channel outputs in 10 column blocks. These 32 outputs read out 32 adjacent pixels in a row at a time. The pixel clock of the detector is set to 10\,MHz, which enables a readout speed of about 640\,Mpixels/s. 32 adjacent pixels are sent every clock tick in such a way that 10 pixel clock ticks are required to read one line of the detector on full frame mode.  This capability increases the frame rates greatly (achieves up to $3500$ frames/s for a full window readout). SAPHIRA also supports sub window readout mode. 

\subsubsection{Operation}

The SAPHIRA detector is placed inside the C-RED ONE camera cryostat with a sealed  vacuum environment. The camera itself has an integrated cooling system with a pulse tube technology. The C-RED ONE operates at $10^{-5}\,$mbar pressure and 80\,K temperature. It images the MIRC-X spectrograph at room temperature through a $f/4$ aperture. The camera has four cold filters close to the detector in order to block the background flux outside the J and H-band wavelength window.

\subsubsection{Readout modes and data acquisition strategies}

The C-RED ONE camera has several standard readout modes (single read mode, correlation double sampling mode, multiple non-destructive reads mode) including a custom requested IOTA readout mode. The IOTA readout mode was used previously for a PICNIC camera at the IOTA array\cite{Pedretti2004}. In this mode, multiple consecutive non-destructive reads are done for the same pixel ($N_\mathrm{reads}\times$) before reading the next pixel. In the same way, the line is read multiple times ($N_\mathrm{loops}\times$) before reading the next line. At the end of a frame, the reads start over again with the first pixel of the window. The detector is reset every $N_\mathrm{fpr}$ frames, with $N_\mathrm{fpr}$ typically in the range 50 to 200 in our application. In practice, we found that the IOTA readout mode provides the best noise performance at our frame rates.
 
The maximum IOTA readout frame frequency varies as below:
\begin{equation}
FPS =  \dfrac{10\mathrm{~MHz}} { N_\mathrm{reads}} \dfrac{1 }{ (N_\mathrm{x}/32  + N_\mathrm{loops} + 3/N_\mathrm{reads} + N_\mathrm{loops} -1 ) \times ( N_\mathrm{y} + 3) }
\end{equation}

Where, $N_\mathrm{x}$ and $N_\mathrm{y}$ are the user-specified width and height of the detector, in pixels, respectively. $N_\mathrm{reads}$ and $N_\mathrm{loops}$ are the number of times a pixel and a line are read, respectively.
 
SAPHIRA arrays have been used for wavefront sensing and fringe tracking applications (Gravity Collaboration et al. 2017\cite{GRAVITY2017} and references within). However, to our knowledge, it is the first time it is used as the primary scientific camera in an astronomical instrument.

\subsection{Image acquisition and computer}

The C-RED ONE image acquisition server handles the image acquisition of the camera. Two SDR-26 camera link cables are used to grab the frames from the camera firmware to the MIRC-X data acquisition computer. One camera link cable communicates through the serial line (115200 Bauds, 8 bits, No parity and 1 stop bit) embedded in it. This is used to alter the configuration of the camera. The second communication cable is used for frame grabbing. 

On the MIRC-X computer side, frames are grabbed by a Matrox Radient eV-CL frame grabber installed on the PCIe 2.0 $\times$ 16 slot. It allows a data speed of peak bandwidth of up to 4GB/s with a pixel clock speed of 80\,MHz. This speed is sufficient as the camera produces 540 MB/s data at maximum ($320 \times 256~{\rm pixels}^2  \times 16{\rm -bit} \times 3500$ frames/s).

\begin{figure} [h!]
  \begin{center}
    \begin{tabular}{c}
    	\includegraphics[width=6.5in]{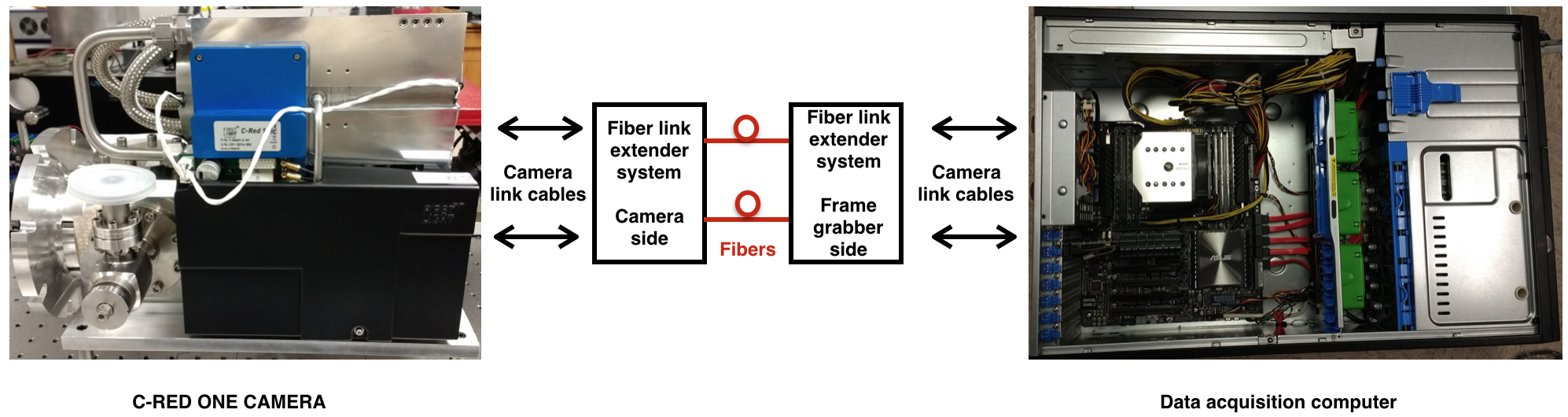}\\
        \includegraphics[width=6in]{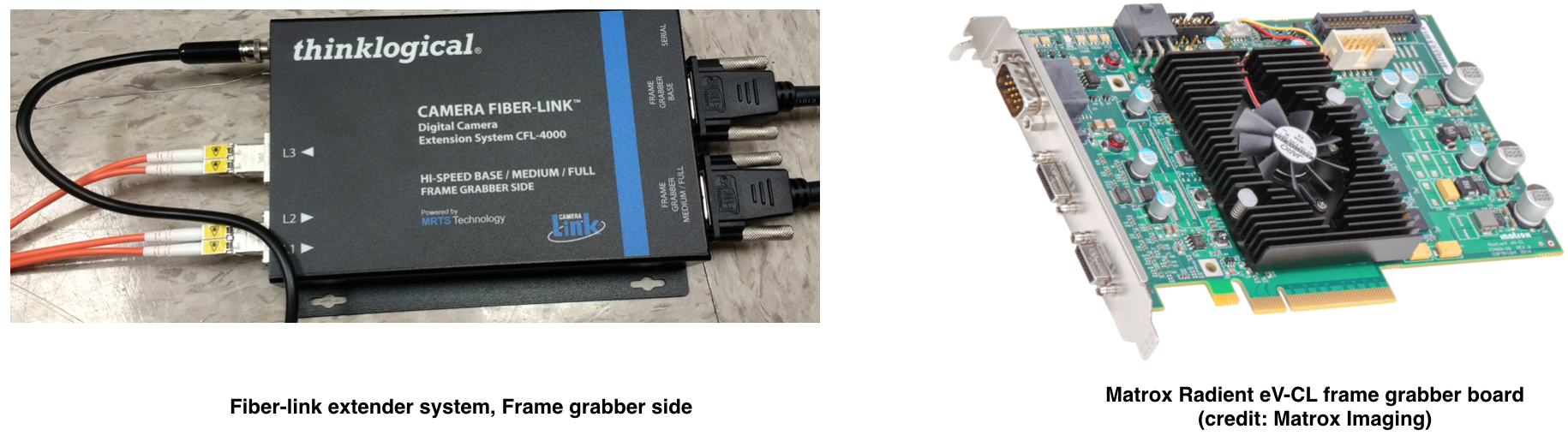}
    \end{tabular}
  \end{center}
  \caption[example] 
	  { \label{fig:Camera_fiberlink} 
	    A camera-fiber-link extender system, CFL-4000, from thinklogical  is used to connect the C-RED ONE camera in the beam combiner lab to the Matrox frame grabber installed on the data acquisition computer outside the lab.  
	  }
\end{figure}

The camera (inside the beam combiner lab) and the data acquisition computer (placed outside the beam combiner lab) are separated by a few tens of meters. For this reason, the camera-fiber-link extender system from thinklogical is used. The camera-fiber-link extender system (Figure~\ref{fig:Camera_fiberlink}) uses a camera side unit and a frame grabber or computer side unit that are connected by duplex multi-mode fiber optic cables from Fibertronics Inc. In theory, this system allows camera link video transmission over up to 350 meters with no loss of signal and without the use of amplifiers.

The MIRC-X computer embarks a low latency 2TB SSD disk in order to write the images at the necessary speed. This disk is a Samsung 960 Pro, mounted on the M.2 port. The computer runs Xubuntu 16.04 operating system on a dedicated 256 GB SSD. At the end of each night, the MIRC-X FITS data is compressed using \textsl{fpack} software and transferred to two 8TB HDD disks. 

\subsection{Vacuum system}

The C-RED ONE camera requires $\le 10^{-5}$ mbar vacuum pressure for the operation. The camera has four cold long-wavelengths blocking filters. Two of the them are placed very close to the SAPHIRA detector and because of them, vacuum is trapped between the filters and the detector. A sudden change in vacuum pressure could break the filters and thus requires protection against this occurring.

The required vacuum is achieved by using a HiPace 80 turbopump from Pfeiffer (achieves $\le 10^{-6}$ mbar). Once the camera is cooled down and during on-sky observations, the vacuum is maintained by an ion pump from Gamma vacuum (SPC PN 900026) for the following reasons: (1) to maintain low-vibrations; (2) for security against power outages (the turbopump draws a lot of power and is not on UPS system).  The ion pump has proved to be sufficient to maintain the required vacuum (keeps $\le 10^{-7}$) for several days when the camera is kept cold). The turbopump is mainly used to pump the camera before cool down, during cool down, and during warming up.

\section{Software}
\subsection{High level specifications}
High level specifications of the MIRC-X software are summarized below:

\begin{itemize}
\item Image acquisition should not miss frames even at maximum frame rate 3500 frames/sec in the full window mode. 

\item MIRC-X should provide fringe tracking with low latency (specification 50\,ms for group delay tracking; goal 5\,ms for phase tracking). 

\item MIRC-X should work simultaneously, that is to control both the CHARA delay lines and some internal delay lines to keep both MIRC-X and MYSTIC instruments (science and fringe tracker) within the coherence length.

\item MIRC-X should record data simultaneously, with at least 95\% of overlap in time (no more than 5\% of the time one instrument shall be integrating while the other is not). This is necessary to ensure efficient a-posteriori fringe tracking in the pipeline. Consequently, a common data sequencer is required to synchronize the MIRC-X and MYSTIC data taking.  

\item MIRC-X should control all motorized actuators or motors in the warm optics and inside the cryostat.

\item Polarization explorer enables the optimization of the maximum visibility by aligning the polarization plates. It should automate the optimization of the polarization plate position based on the observed visibility and/or the observed polar-differential phase.

\end{itemize}

\begin{figure} [h!]
	\begin{center}
		\begin{tabular}{c} 
			\includegraphics[width=4in]{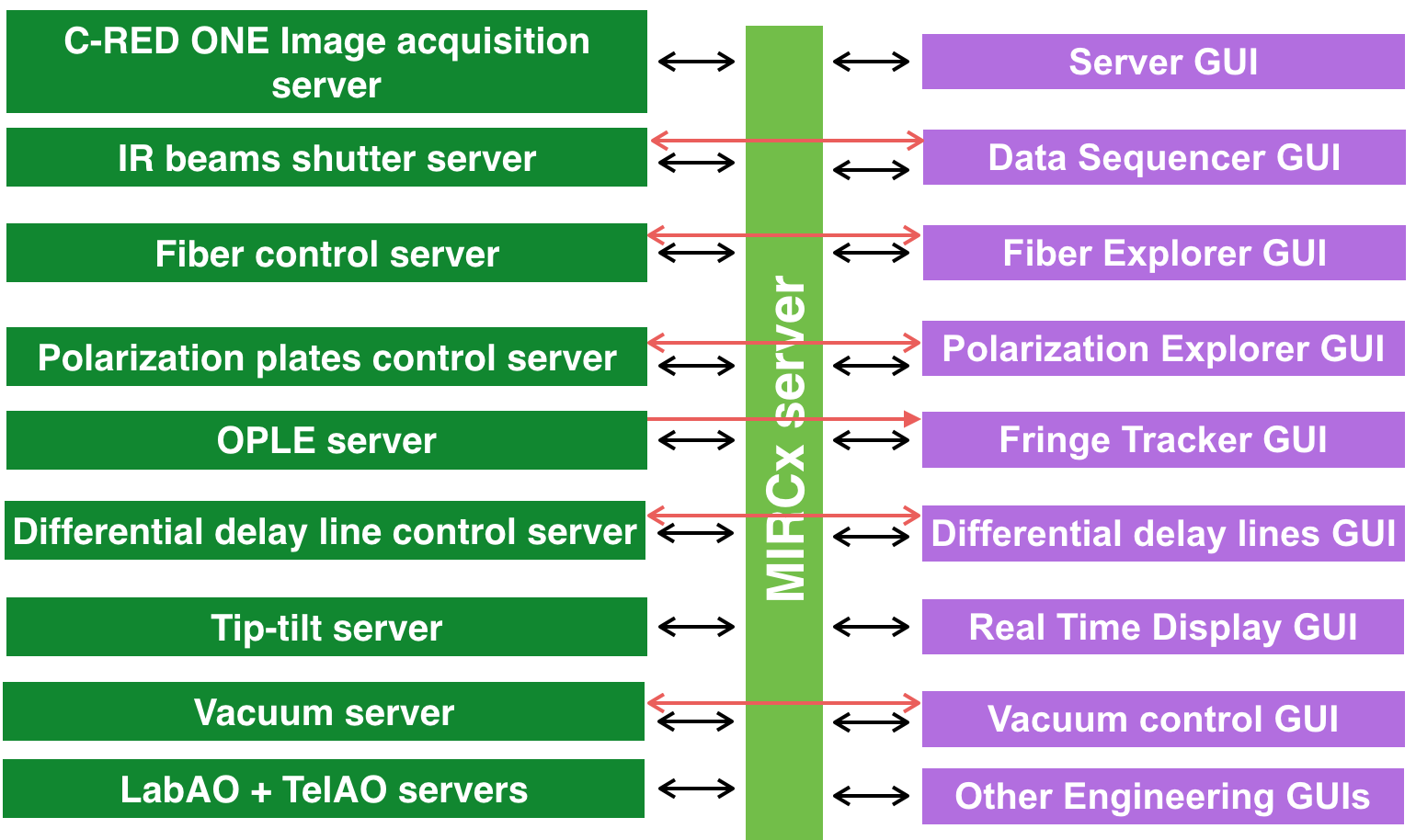}
		\end{tabular}
	\end{center}
	\caption[example] 
	{ \label{fig:software} 
		The MIRC-X instrument server communication with the various servers and GUIs.  The arrow marks indicate the communication.
	}
\end{figure}

\subsection{Architecture}

All the MIRC-X software runs on a Linux operating system. The MIRC-X software follows the CHARA standard ``client/server" model and runs under Linux. The hardware such as camera, motors/actuators and sensors are controlled by C-written servers. A server is a program that accepts requests, performs a task and returns some data. The servers can then have multiple clients (GUIs, or other servers) that make requests to them.  All the MIRC-X servers run on the MIRC-X computer. Many other servers run on different machines inside CHARA, and are accessible via TCP socket message protocol (e.g shutter server, delay line serve, adaptive optics servers). 

The GUI clients are GTK-based (Gimp Tool Kit). GTK is a software library for developing X windows programs. Scientific plots inside the GUIs are made use PLplot. The GUI can be running on various machines, locally or remotely. They connect to the servers via the standard CHARA protocols. Figure~\ref{fig:software} presents the list of servers and clients that will be used to operate the MIRC-X instrument. 

The software includes the control of the data acquisition to grab, process and save the camera images, different motorized functions and sensors, and the GUIs.

\subsection{MIRC-X Data Acquisition System}

\begin{figure} [h!]
  \begin{center}
    \begin{tabular}{c} 
      \includegraphics[width=5in]{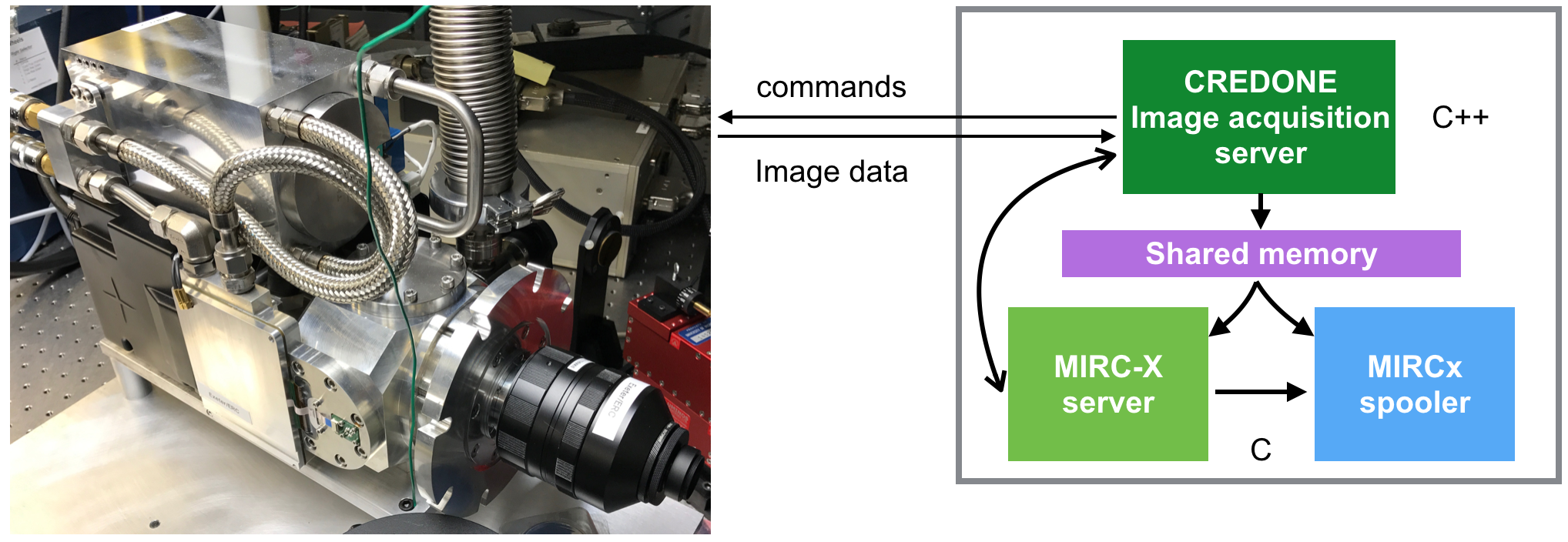}
    \end{tabular}
  \end{center}
  \caption[example] 
	  { \label{fig:daq} 
	    MIRC-X camera data acquisition software.  The C-RED ONE camera images are acquired to the MIRC-X computer using a Matrox frame grabber and written to a shared memory. The MIRC-X instrument server reads the shared memory and does the image processing and fringe tracking. 
	  }
\end{figure}

First Light Imaging company delivered us a demo data acquisition software and that is improved and adapted for our case. The challenge was to read 3500 frames/s and write 540 MB/s of data with a minimum latency. The camera data acquisition is handled by an in-built C-RED ONE image acquisition server (Figure~\ref{fig:daq}).  This program grabs the images from the camera to the MIRC-X computer and writes them to a circular buffer-based shared memory without losing frames (i.e., 3500 frames per seconds) and with a minimum  latency (time lag between two frames: mean 0.2\,ms and maximum 0.5\,ms). The software is written in C++ using Matrox Imaging Library (MIL) version 10.  The MIRC-X instrument server reads the images from the shared memory and does pre-processing, power spectrum computation and fringe tracking. The MIRC-X instrument server communicates with the C-RED One image acquisition server to modify the camera configuration upon request from the GUIs. The MIRC-X spooler is the data saving process. It reads images from the shared memory and writes to FITS files. 

We ignore the first read in the ramp after the reset because it has spurious flux in it. This may be caused by settling effects immediately following the reset. We take the mean of $N_\mathrm{reads}$ and $N_\mathrm{loops}$ frames and then compute the difference of two frames (i.e., correlation double sampling) depending on coherent integration exposure requirement which removes the $kTC$-reset noise.

\subsection{CHARA-compliant servers and GUIs}

During the first few months of MIRC-X operation, it was operated with Python GUIs which were taken from MIRC and modified for the MIRC-X detector. However, we developed CHARA-compliant servers and GTK-based GUIs for the following reasons:

\begin{itemize}
\item Python GUI software reached a position where further development would complicate things because of its design. Python GUIs communicate C-structures with the MIRC-X servers and these structures are very system-dependent for Python.
\item CHARA-compliant software development was critical for the co-phasing operation of MIRC-X and MYSTIC. 
\item Remote observing capability requires servers and GUIs in the same programming language (here C).
\item Easy to communicate with CHARA servers. Easy to support and maintain by the CHARA team.
\end{itemize}
	
The actuators or motors used for MIRC-X are from different companies and they accept different protocol commands to move to a position, to retrieve a position, or to home the actuator. In order to unify the command system, for each device, a server program is made so that it simplifies the command system and logs the positions of actuators. The actuator or motor controllers are connected to an MOXA NPort device using a DB9-to-RJ45 conversion connector and commanded using the TCP protocol.

Below, all the servers involved in the operation of the MIRC-X instrument are reported. The servers that run both on MIRC-X and MYSTIC actually share the same source code, and are just started with different runtime options.

\begin{figure} [h!]
	\begin{center}
		\includegraphics[width=6in]{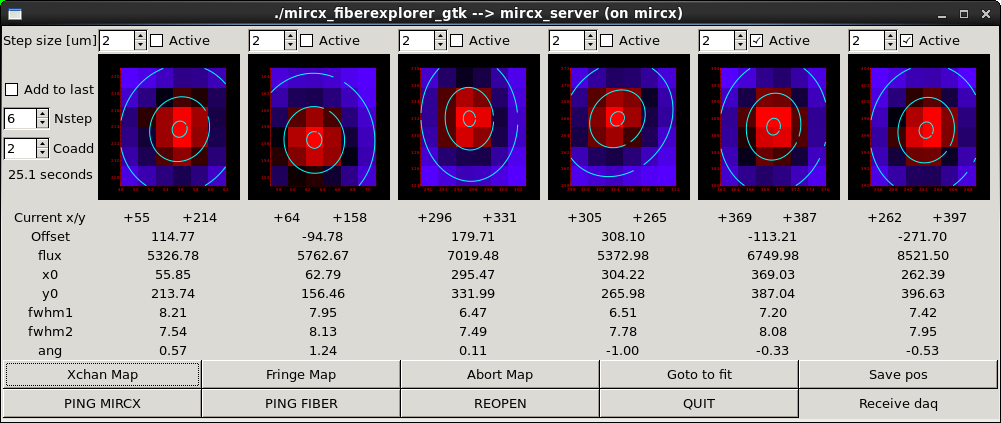}
	\end{center}
	\caption[example] 
	{\label{fig:fiber_explorer}  
		The fiber flux explorer GUI. The fiber explorer step size (in microns) and number of steps can be changed. The six image windows correspond to the six telescope beams. The contour plots show the result of the Gaussian fit. The Gaussian fit results can be seen.
	}
\end{figure}

\begin{itemize}
\item \textbf{Instrument sever} is the heart of the instrument. It configures the camera. It reads images from the shared memory, optimizes flux into fibers by communicating with the fiber server, does image pre-processing, computes power spectrum and group-delay offset, and fringe tracks by sending the offsets to the CHARA delay line server, or the differential delayline server.
  
\item \textbf{Fiber server} controls the fiber motors. It enables optimization of the flux injection into the fibers. A fiber flux exploration is carried out by moving fibers searching for flux. The resulting map is fitted with a Gaussian model to find the position that results in the best flux injection. The map is updated in real-time on a dedicated GUI (Figure~\ref{fig:fiber_explorer}).

\item \textbf{Polarisation server} controls the polarization plates to optimize the visibility contrast (see Section \ref{sec:polarization} for more details). 
  
\item \textbf{Differential delayline server} controls the Zaber motors (T-LA28A) to move the internal differential delay line mirrors. These are used to adjust the optical OPD to align MIRC-X in-phase with MYSTIC and other CHARA combiners ($\sim 10$\,mm). It provides 20\,mm-mechanical range with a step size $0.1\mu$m and repeatability $<0.4\mu$m.
  
\item \textbf{Vacuum server} is an engineering-based server which controls the electromagnetic valve located between turbopump and the ion pump using an iBoot device based on the pressure threshold. It also reads pressures from turbopump Pfeiffer gauge and ion pump gauges (SPC PN 900026) and logs them with time stamps.
\end{itemize}

\begin{figure} [h!]
	\begin{center}
		\begin{tabular}{c} 
			\includegraphics[width=6in]{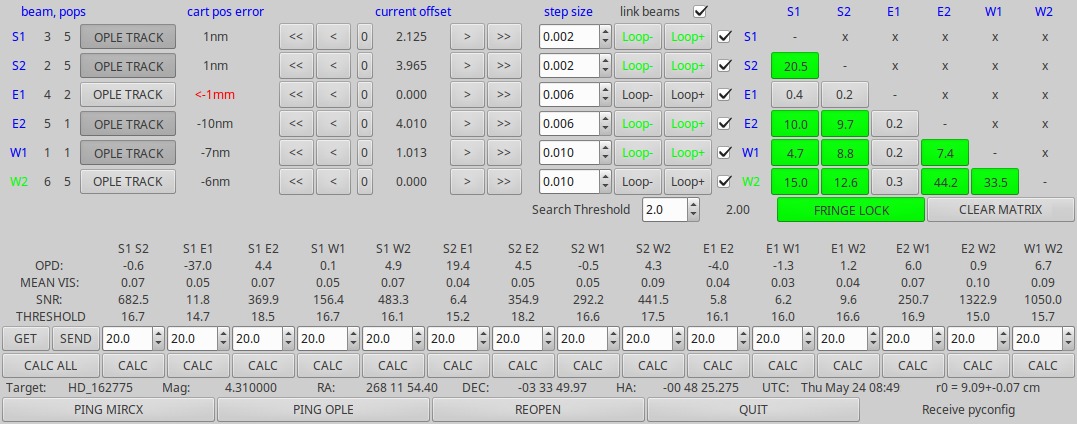}
		\end{tabular}
	\end{center}
	\caption[example] 
	{ \label{fig:GDT_GUI} 
		The group delay tracking GUI, written in GTK.   The Loop-/Loop+ buttons are used to search for the fringes. Once the fringes are found the matrix buttons (in green color) are used to lock the fringes. Delayline carts tracking errors and MIRC-X offsets with respect to the baseline model can be seen. At the time this photo was taken, the E1 telescope was not available because of re-coating. 
	}
\end{figure}

\begin{figure} [h!]
  \begin{center}
    \includegraphics[width=3.5in]{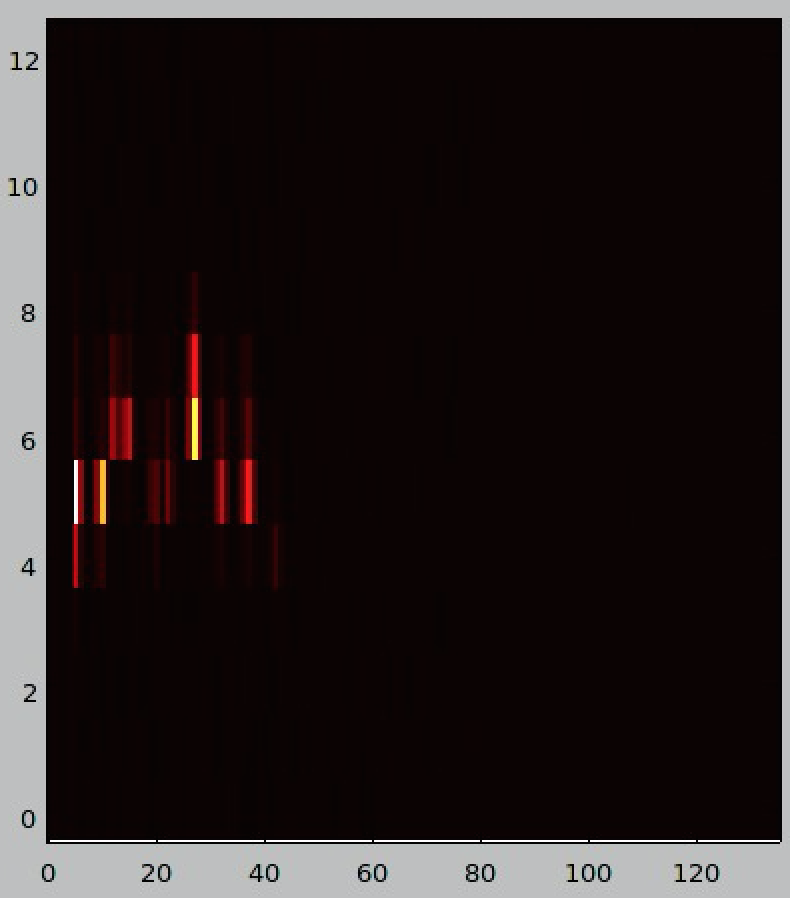}
    \includegraphics[width=3in]{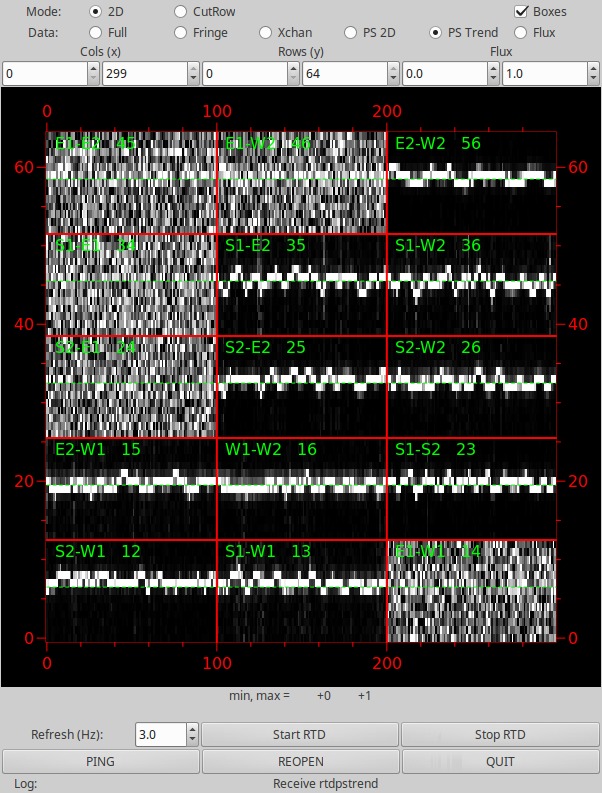}
  \end{center}
  \caption[example] 
	  {\label{fig:RTD_GUI}  
	    The left panel shows peaks of power spectrum of fringes (10 baselines) resulting from the combination of five telescope beams in a non-redundant baseline distribution. The right panel shows OPD scans (waterfall display). These displays are updated in real-time. At the time this photo was taken, the E1 telescope was not available due to re-coating.  
	  }
\end{figure}

Observing with the MIRC-X system requires starting servers in a  sequence as some servers open writing shared memory and others started reading the memory. We use the same server and GUI executables for MIRC-X and MYSTIC with a runtime option.  This allows us easy maintenance of code and reduces complexity of operation. Three new GTK-written fiber explorer, group delay tracking and waterfall GUIs are highlighted here (Figure~\ref{fig:fiber_explorer}, \ref{fig:GDT_GUI} and \ref{fig:RTD_GUI}). 

\subsection{Fringe tracking}
During observing or fringe searching, the MIRC-X instrument server computes the fringe tracking SNR and group delay for all fringes in each incoming frame, accounting for the number of coherent and incoherent integrations specified by the user. The fringe loop (i.e.\ search) is initiated by moving the CHARA delay lines in regular steps until the SNR reaches the required threshold. Then fringes are tracked by sending the measured OPD to the CHARA delay lines. The algorithm uses the closing triangles to lock fringes on more baselines. The tracking algorithm runs entirely on the MIRC-X instrument server and the user interacts with it via a dedicated GUI (see Figure~\ref{fig:GDT_GUI} ). 

The simultaneous operation of MIRC-X and MYSTIC will have to deal with the CHARA delay lines, to track the turbulence, and with the internal delay line of at least one of the two combiner, to track the instrumental drift and atmospheric dispersion. We are exploring two possible schemes:

\paragraph{Master - slave scheme} In this approach, one beam combiner is defined as “master” and controls the CHARA delay lines. The second beam combiner is used as “slave” and it controls the differential delay lines. This could be rather simple to implement as we would only need to change what actuators the GUIs control. This scheme would be efficient in the situation where one combiner (master) has systematically a higher SNR than the other (slave), for whatever instrumental or astrophysical reason.

\paragraph{Combined scheme} For very resolved object, where baselines might reach 0 visibility depending on their length and wavelength, the previous scheme is non-optimal. Ideally, we want to combine all the available information, and decide in real time which delaylines to correct either the CHARA and/or the differential. 

\subsection{System monitoring and logging} 
\begin{figure} [h!]
  \begin{center}
    \includegraphics[width=4in]{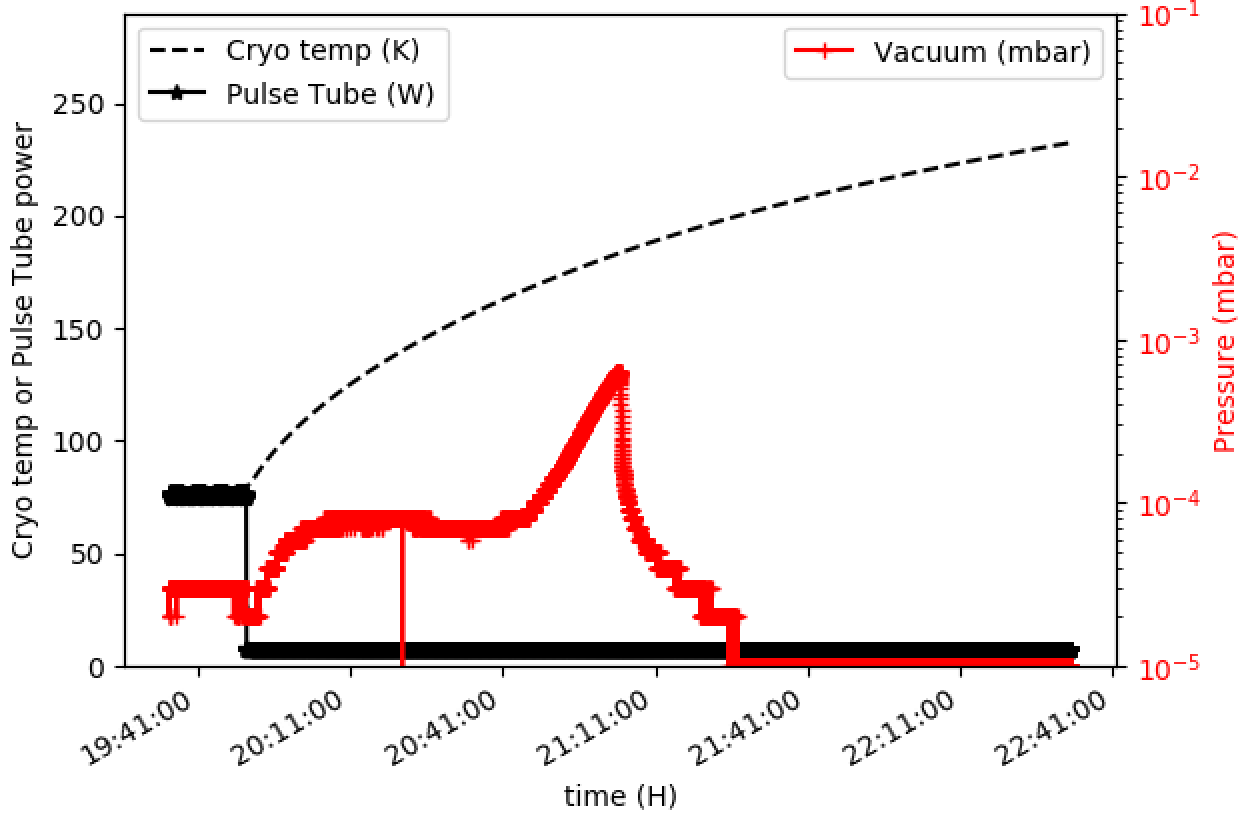}
  \end{center}
  \caption[example] 
	  {\label{fig:mircx_warming}  
	    Vacuum pressure, cryo cooler temperature and pulse tube power consumption during a typical camera warm-up cycle. The power consumption drop of pulse tube indicates the start of the warm-up procedure. The increased pressure while warming up is due to out-gassing. 
	  }
\end{figure}

A lot of effort has gone into automating the operation of MIRC-X.  Also, in order to monitor the health of the instrument, various parameters of the camera and the instrument such as the temperature of various sensors; camera status; the pressure of the camera, turbopump, and ion pump are all logged to disk every second. Real-time plots (ex. Figure~\ref{fig:mircx_warming}) are generated for easy monitoring of the parameters. Using these parameters, automatic alert emails are generated and sent to the engineering team when something suspicious occurs with the instrument.  

\subsection{Sequencer}
The purpose of the sequencer is to acquire a sequence of files, with specific file types (e.g.\ data with a certain shutter open, background data, file with fringe data, ...). Each data set contains several individual files and are numbered incrementally.  Currently, the user specifies how much files are to be recorded for each data type. Recording one data file typically requires only a few seconds, depending on various parameters of the data acquisition system. 

This strategy has to be modified to allow the operation with both MIRC-X and MYSTIC. Indeed, different individual exposure times may be required for the two beam combiners based on the magnitude of a target. The sequencer GUI will be modified so that the user will enter a total integration time, such as 30s. The sequencer GUI will convert this common integration time into a specific number of data files for each instrument. Indeed, the recording time for a file can be predicted accurately (within 1/300) from the data acquisition parameters.






\section{Installation of polarization controller}\label{sec:polarization}
\begin{figure} [ht]
  \begin{center}
    \begin{tabular}{c} 
      \includegraphics[width=6in]{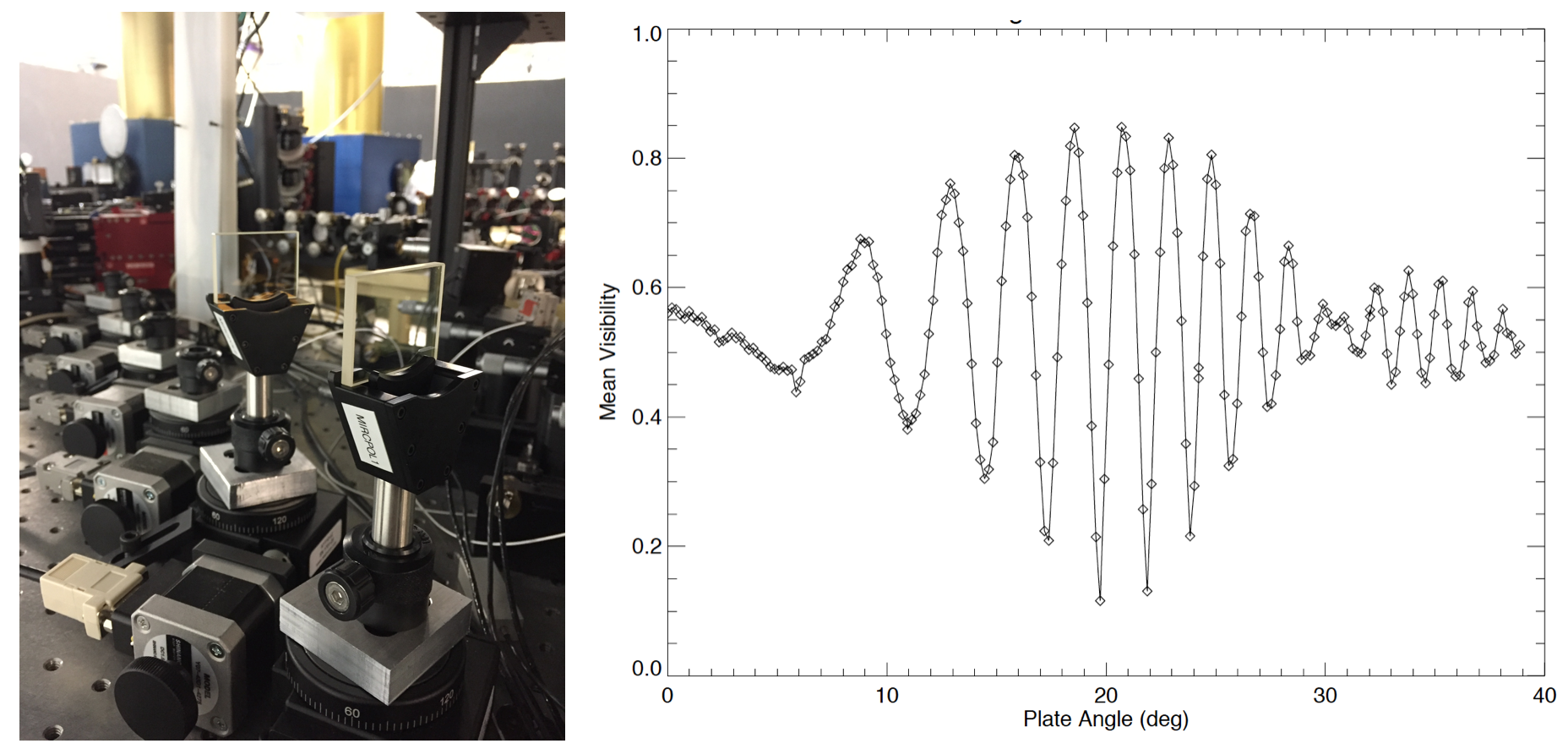}
    \end{tabular}
  \end{center}
  \caption[example] 
	  { \label{fig:polarization4} 
	    Picture of the MIRC-X polarization plates (left) and wavelength-averaged fringe visibility versus the plate angle (right).  
	  }
\end{figure}

Polarization plates (Lithium Niobate) were commissioned in May 2017 and they enable the optimization of maximum visibility for MIRC-X by aligning the polarization plates.  By rotating these plates, an adjustable phase-shift between the vertical/horizontal axes (birefringence axes of fibers) can be introduced, as described in Lazareff et al\cite{lazareff2012}. We carried out an experiment with MIRC, measuring visibility as a function of plate rotation angle (see Figure~\ref{fig:polarization4}). This experiment demonstrated that this design can modulate more than 5 fringes around the mean angle of 20\,deg, with a resolution better than $\lambda$/20.  

These plates have 4\,mm thickness, anti reflection-coating and z-cuts. They have $>$98\,\% transmission over the H-band for incidence angles from 0 to 30 deg, and a transmitted wavefront distortion $<$100\,nm. The polarization plates are mounted on stepper motors, which are controlled by the R256 micro-stepping driver from Lin Engineering and with a CHARA-based controller.

The following two strategies of polarization exploring are contemplated: (1) to measure the visibility in natural light; (2) to measure the phase-shift in polarization split mode.

The visibility calibration is expected to be repeated from time to time, but it is unknown how frequent this will need to be carried out.  Ideally, we want to be able to do it every day, at the start-up. The main limitation is the fact that the CHARA internal light allows forming fringes on one pair of beams at a time only, making the alignment process very slow and tedious.

\section{CHARA laboratory beams characterization}
The characterization of the C-RED ONE camera readout noise and background noise as a function of avalanche gain and frame rate is presented in detail elsewhere in these proceedings (see Lanthermann et al.). Sub-electron readout noise is achieved when avalanche gain  $> 20$. See Table~\ref{table:2} for the background current per second of the camera.  The camera is optimal for avalanche gain between 10 and 100. For gain $> 100$, we get Trap Assisted Tunneling caused dark current. We are background limited for 300\,Hz frame rate or less.

\begin{table}[h!]
  \caption{Background currents per second of the C-RED ONE camera.}
  \label{table:2}
  \centering
  \begin{tabular}{ |c|c|c|c|c|} 
    \hline
    $N_\mathrm{reads}$ & 	$N_\mathrm{loops}$  & Gain=1 & Gain=10  & Gain=100 \\ 
    \hline
    4  &        2      &    200.3  &   59.6      & 171.0 \\
    8   &        2         &  153.1    & 54.4  & 164.4\\
    16   &        2           & 122.5   & 51.3    & 153.9\\
    32   &        2           & 92.1   & 49.1   & 138.5\\
    \hline
  \end{tabular}
\end{table}

The C-RED ONE camera suffers from a radio frequency interference noise (with a frequency of about 90\,Hz) and it is the dominant noise source when operating with the lower gains and fainter signals. We can minimize its effects in the following ways: (1) since the radio frequency interference noise is independent of avalanche gain, one can minimize its effects by amplifying the signal of photons with a higher avalanche gain; or (2) by computing a mean of non-illuminating  pixels in the edges in each row of the detector and then subtracting the computed  mean from all the  pixels in that row. 

The performance of MIRC-X has been characterized using CHARA beam combiner lab by obtaining two-beam fringes. The results are presented in the following sections. 

\subsection{Comparison with the previous PICNIC detector}

\begin{figure} [h!]
	\begin{center}
		\includegraphics[width=6in]{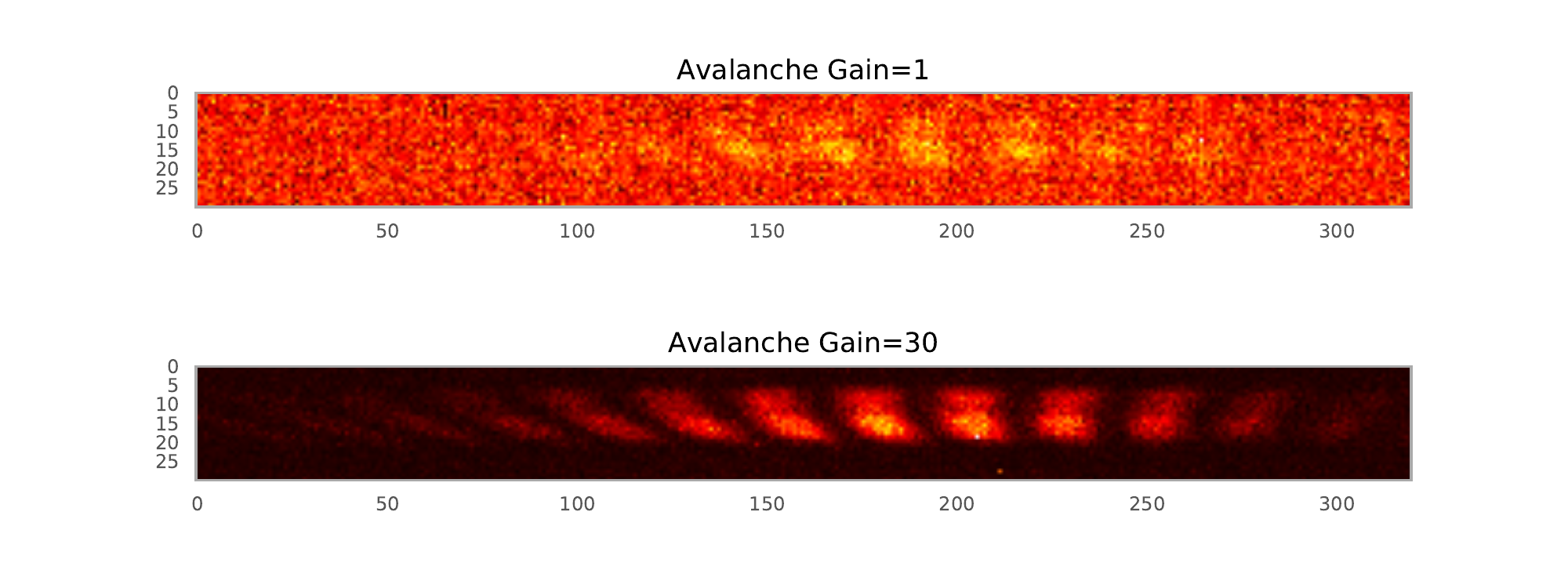}
	\end{center}
	\caption[example] 
	{\label{fig:image} Two-beam fringes obtained with a white light source in the laboratory at gain one (above) and at e-APD avalanche gain 30 (above). 
	}
\end{figure}

We refer to Figure~\ref{fig:image} for laboratory fringes at avalanche gain 1 (approximately equivalent to previous MIRC PICNIC detector) and at gain 30. It shows an improvement of signal-to-noise-ratio (SNR) of fringes when higher avalanche gain is applied. This SNR is improved because the effective readout noise decreases with the avalanche gain ($\sim N_\mathrm{RON}/\mathrm{gain}$).

\begin{figure} [h!]
	\begin{center}
		\includegraphics[width=4in]{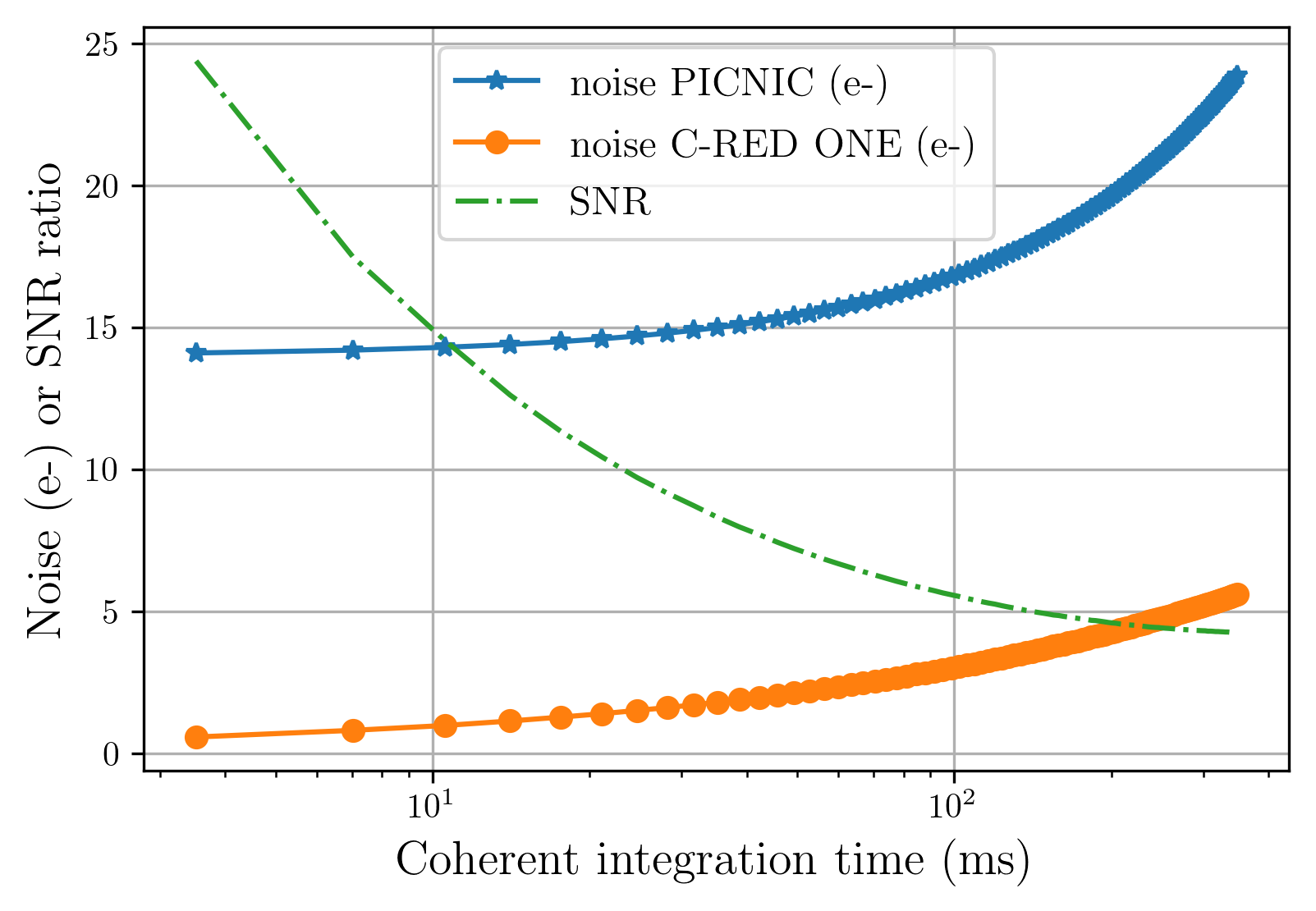}
	\end{center}
	\caption[example] 
	{\label{fig:line_cut}A comparison of the PICNIC with C-RED ONE camera in terms of total noise and SNR signal.
	}
\end{figure}

  The MIRC camera has a noise of $N_\mathrm{RON}=14~e^-$/read/pixel and with a background of 340\,$e^-$/s/pixel whereas the C-RED ONE camera has noise of $N_\mathrm{RON} \le 1~e^-$/read/pixel at avalanche gain of 60 with a background of $\sim 90$ $e^-$/s/pixel. We compute the C-RED ONE camera will have better SNRs with factors of 14 and 5.5 in comparison to the MIRC camera at 10\,ms and 100\,ms coherent exposures (Figure~\ref{fig:line_cut}).

\subsection{Choice of avalanche gain}\label{sec:GAIN_SNR}

The MIRC-X performance is characterized as a function of avalanche gain by obtaining CHARA laboratory two-beam fringes.  We tuned the flux of beams such that MIRC-X gets fringes with a low SNR of fringes ($SNR_\mathrm{fringe}$)  at unity avalanche gain. Then data sets (fringe data, background data and beams data) are taken by increasing the avalanche gain and keeping all other MIRC-X configuration unchanged. Next, the data is reduced with the \footnote{https://gitlab.chara.gsu.edu/lebouquj/mircx\_pipeline.git}{mircx pipeline} authored by J.-B. Le Bouquin. The radio interference noise is cleaned by computing a mean of non-illuminating pixels in the edges in each row of the detector and then subtracting the computed  mean from all the  pixels in that row.  The $SNR_\mathrm{fringe}$ is computed as the ratio of maximum of the power spectrum peak in fringe data to the maximum of the power spectrum peak in the background data.

\begin{figure} [h!]
  \begin{center}
    \includegraphics[width=6in]{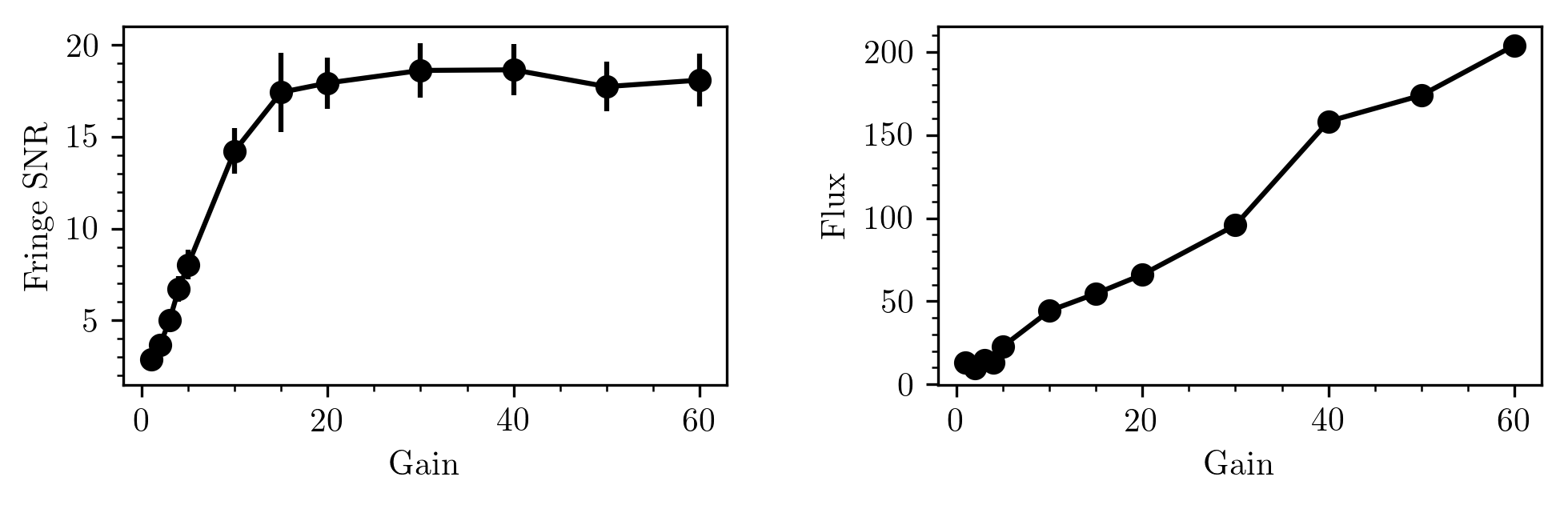}
  \end{center}
  \caption[example] 
	  {\label{fig:Gain_vs_SNR} 
	    MIRC-X performance as a function of avalanche gain: (left) SNR of fringe detection. (right) A photometric channel flux. These plots are obtained from laboratory (no atmospheric turbulence) two beam combiner fringe data.
	  }
\end{figure}

\begin{figure} [h!]
  \begin{center}
    \includegraphics[width=6in]{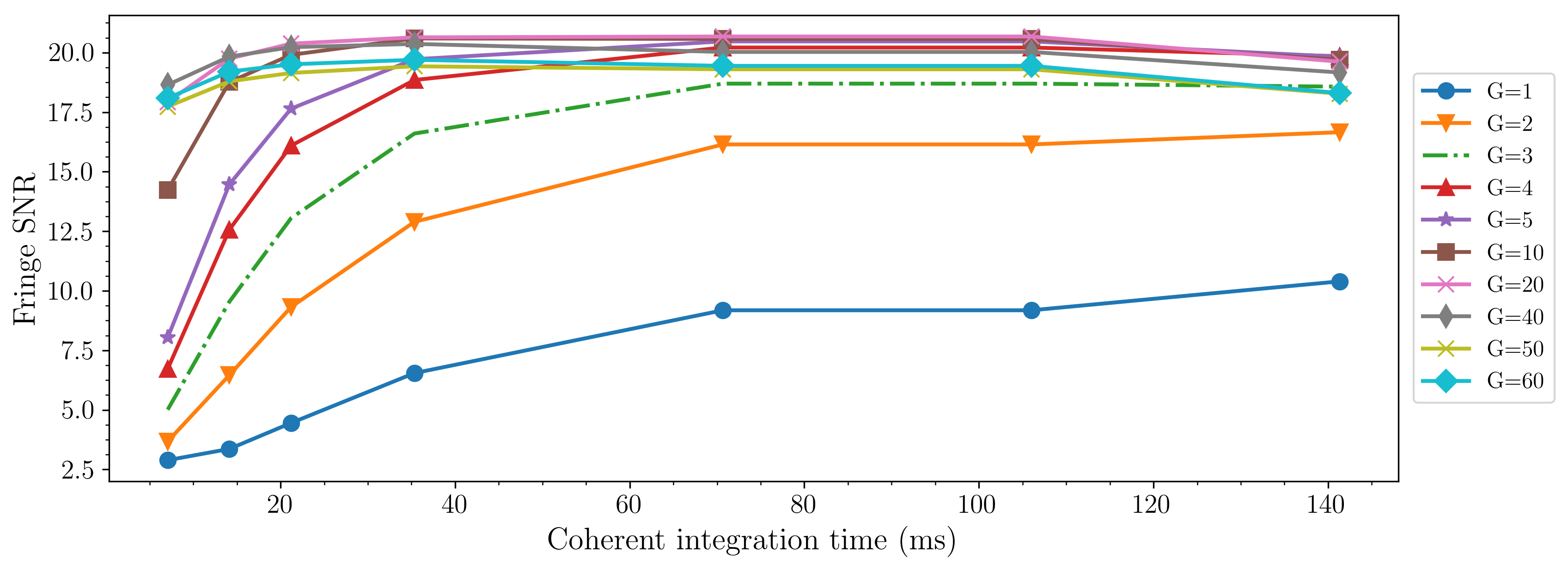}
  \end{center}
  \caption[example] 
	  {\label{fig:Gain_vs_Coherent} 
	    MIRC-X fringe SNR performance as a function of avalanche gain and coherent integration time. These plots are obtained from laboratory (no atmospheric turbulence) two beam combiner fringe data.
	  }
\end{figure}

Figure~\ref{fig:Gain_vs_SNR} presents the $SNR_\mathrm{fringe}$ as a function of avalanche gain at 7\,ms coherent integration time.  The SNR is increased with avalanche gain and flattened when the photon-limit  ($N_\mathrm{Photons} > N^2_\mathrm{RON}$) is reached. Here, $N_\mathrm{Photons}$ and $N_\mathrm{RON}$ are the number of photons (in electrons/s/pixel) and number of readout noise (in electrons/pixel).

Figure~\ref{fig:Gain_vs_Coherent} presents the $SNR_\mathrm{fringe}$ as a function of coherent integration time for various avalanche gains.  An avalanche gain around $40$ and a shorter coherent   exposure are the best choices for MIRC-X in the case of bad seeing. For good seeing, the coherent integration time can be increased depending on the atmospheric coherence time.

For a comparison, the MIRC PICNIC camera performance is approximately equal to unity avalanche gain performance of the MIRC-X camera. By using an avalanche gain of 40, the $SNR_\mathrm{fringe}$ is increased by a factor of 7 ($\sim2$ magnitude additional sensitivity) in comparison to unity avalanche gain. This is for a  coherent integration of 7\,ms. For 100\,ms coherent integration,  the $SNR_\mathrm{fringe}$  improvement is only a factor of 2. However, the laboratory fringes data we took did not cover the higher readout-noise-limited regime (i.e., $N_\mathrm{Photons} << N^2_\mathrm{RON}$). If that is the case, we would expect even more $SNR_\mathrm{fringe}$  improvement.  We have not realized the anticipated SNR performance (a factor of 17 at 7\,ms coherent integration, see Figure~\ref{fig:Gain_vs_SNR}). For this upgrade of optics is planned in Phase 2 to address the over sampling of fringes, which is caused by the pixel scale difference between the old PICNIC (40~$\mu$m) and the new C-RED ONE (24~$\mu$m) cameras.

\subsection{Choice of $N_\mathrm{reads}$ and $N_\mathrm{loops}$ of the IOTA readout mode}
By increasing the $M=N_\mathrm{reads} \times N_\mathrm{loops}$, the detector readout noise is expected to reduce by a factor of $\sqrt{M}$ \cite{Mclean2008}, i.e.\ $ N^\mathrm{eff}_\mathrm{RON} \sim \tfrac{N_\mathrm{RON} } { \sqrt{M}}$. 

\begin{table}[h!]
	\caption{$SNR_\mathrm{fringe}$, beam flux and frame rate as a function of $N_\mathrm{reads}$ and $N_\mathrm{loops}$.}
	\label{table:1}
	\centering
	\begin{tabular}{ |c|c|c|c|c|} 
		\hline
		$N_\mathrm{reads}$ & 	$N_\mathrm{loops}$  & $SNR_\mathrm{fringe}$ & Flux  & Frame rate (Hz) \\ 
		\hline
		4   &        1       &    3.2   &   8       & 3277 \\
		4   &        2         &  14.8    & 15.8   & 1666\\
		8   &        2           & 29.6   & 28.3    & 838\\
		16  &        2         &   55.6     & 66.9   & 425\\
		16  &        4          &  91.9    & 129.21 &  215   \\
		\hline
	\end{tabular}
\end{table}

The $SNR_\mathrm{fringe}$ varies with $M$ as given in Eq.~\ref{SNR},

\begin{equation}\label{SNR}
  SNR_\mathrm{fringe} \sim \dfrac{M N_\mathrm{Photons} }{    \sqrt{ { M N_\mathrm{Photons} + \sigma^2/M }}},
\end{equation}

\begin{itemize}
	\item $SNR_\mathrm{fringe} \propto M^{3/2}$ until it reaches the photon-limit regime ($N_\mathrm{Photons} > N^2_\mathrm{RON}$). 
	
	\item $SNR_\mathrm{fringe} \propto M^{1/2}$ in photon-limit regime. 
\end{itemize}

\begin{figure} [h!]
	\begin{center}
		\includegraphics[width=6.5in]{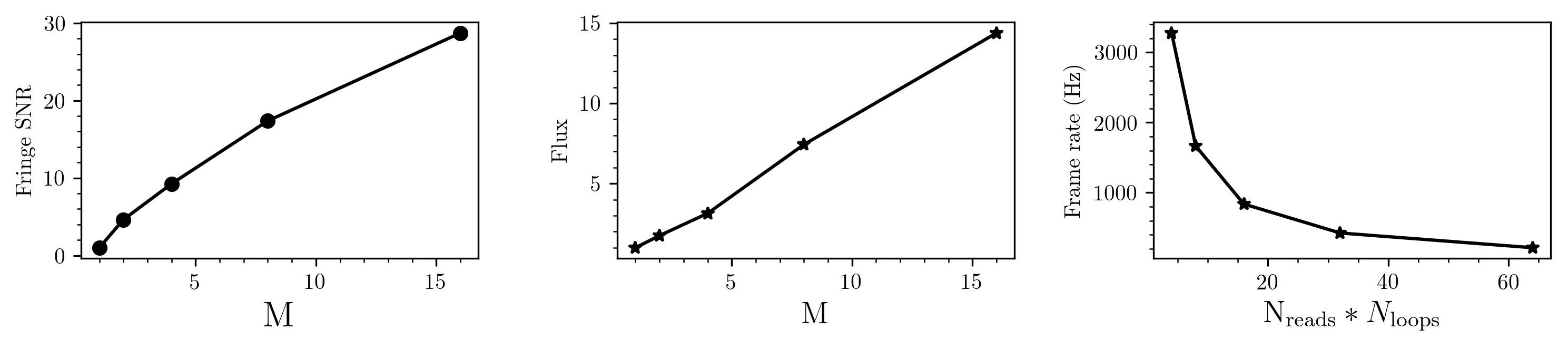}
	\end{center}
	\caption[example] 
	{\label{fig:NREADS_NLOOPS_SNR}  MIRC-X performance as a function of $N_\mathrm{reads} \times N_\mathrm{loops}$ of IOTA readout mode. (Left panel) $SNR_\mathrm{fringe}$,  (middle panel) photometric channel flux of a beam, and (right panel) recorded frame rate (Hz).  Here, the $M=N_\mathrm{reads} \times N_\mathrm{loops}$ normalized with 4.
	}
\end{figure}

 Table~\ref{table:1} presents the fringe SNR measured for various IOTA readout mode configurations.  Figure~\ref{fig:NREADS_NLOOPS_SNR} presents the fringe SNR, beam flux, and frames rates which vary as a function of $N_\mathrm{reads} \times N_\mathrm{loops}$. This data was taken with an avalanche gain of 10. At this gain, the radio frequency interference noise effects are negligible. 

With $N_\mathrm{reads}=16$ and $N_\mathrm{loops}=2$, the IOTA readout mode configuration is optimal for a reduced noise and for shorter coherent exposure time observations.

\section{First light and results}
\begin{figure} [h!]
  \begin{center}
    \includegraphics[width=5.8in]{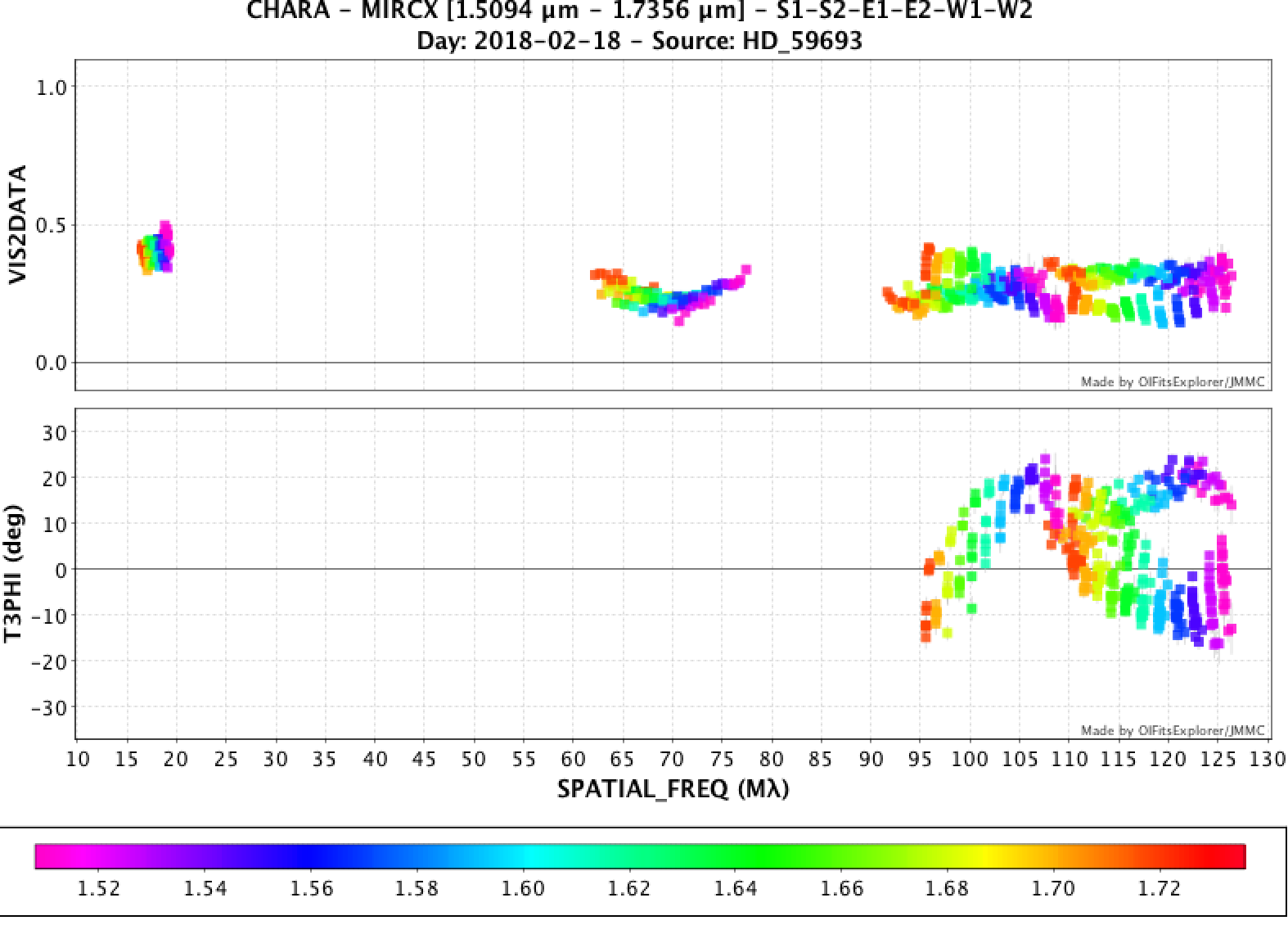}
  \end{center}
  \caption[example] 
	  {\label{fig:VISS2_T3PHI}  
	    An example of calibrated visibility  and closure phase curves observed for 89 Her (5T observation).}
\end{figure}

Commissioning of the MIRC-X instrument was carried out from UT July 03 to UT July 16, 2017.  As of this writing, 68 days of data have been collected in H-band since 2017 July 04 with the majority of observations undertaken for science purposes. So far we have observed more than 5 targets fainter than 6 magnitude in H-band.  Figure~\ref{fig:VISS2_T3PHI} presents visibility square and closure phase measurements for the post-asymptotic giant branch binary object, 89 Her, showing clear features in the closure phase data.

We have yet to quantitatively compare the sensitivity performance of MIRC-X to MIRC. However, this comparison is not easy due to varied throughput of the telescopes and optics, optical alignments, and atmospheric seeing.

\section{Summary and future}
\subsection{Summary}
\begin{itemize}
\item MIRC-X is operated with the IOTA readout mode to achieve a better noise performance. An avalanche gain around $40$ and IOTA readout mode configuration of $N_\mathrm{reads}=16$ and $N_\mathrm{loops}=2$ are the best choices for the MIRC-X operation in the presence of average seeing conditions. 

\item C-RED ONE camera has been commissioned and saw first light in July 2017. The upgrade will improve a factor of more than 15 SNR of fringe detection in comparison to the previous PICNIC detector-based MIRC for short coherent integrations. 

\item Although the SAPHIRA arrays have been used for tip-tilt sensing, wavefront sensing and fringe tracking applications, to our knowledge, this is the first time it has been used as the primary scientific camera in an astronomical instrument.

\item A laboratory-based two beam fringe experiment confirms a boost of SNR with a factor of 7 ($\sim2$ magnitude additional sensitivity) at coherent integration of 7\,ms. However, this test did not cover the higher readout noise-limited regime (i.e., $N_\mathrm{Photons} << N^2_\mathrm{RON}$) where we expect even more SNR improvement. To obtain the anticipated SNR improvement, we are also planning to address the over sampling of fringes, which is caused by the pixel scale difference between the old PICNIC (40~$\mu$m) and the new C-RED ONE (24~$\mu$m) cameras.

\item CHARA-compliant servers and GUIs have been developed and are fully operational. 



\item A prototype of MIRC-X is developed in the Michigan lab to test the Phase 2 optics. 
\end{itemize}

\subsection{Goals for software}
\begin{itemize}
\item Once MIRC-X and MYSTIC are running simultaneously, we will need a more sophisticated fringe tracking algorithm that makes use of all the fringe data from both instruments simultaneously. 
	
\item A polarization explorer strategy will be developed to automate the optimization of the polarization plate position based on the observed visibility and/or the observed polar-differential phase.
	
\item Atmospheric differential refraction introduces {1.5\,arcsec} shift between the visible and infrared photo-centers at Zenith angle of 60\,degrees for average CHARA array conditions. Developing an atmospheric differential refraction correction model may bring an important operational gain because we suspect most of the night-time instability in fiber injection is, in fact, the atmospheric refraction.
	
\item Some GUIs may be merged to save space on monitors and ease operations.
\end{itemize}

\acknowledgments 
MIRC-X is funded, in parts, by a Starting Grant from the European Research Council (ERC; grant agreement No. 639889, PI: Kraus) and builds on earlier investments from the University of Michigan and the National Science Foundation (NSF, PI: Monnier). This research has made use of the Jean-Marie Mariotti Center \texttt{OIFits Explorer} service (http://www.jmmc.fr/oifitsexplorer).

\bibliography{main} 
\bibliographystyle{spiebib} 

\end{document}